\documentclass[reprint, superscriptaddress, longbibliography, nofootinbib]{revtex4-2}

\usepackage{epstopdf}
\usepackage{graphicx}
\usepackage{amsfonts}
\usepackage{amssymb}
\usepackage{amsthm}
\usepackage{array}
\usepackage{amsmath}
\usepackage{verbatim} 

\usepackage[colorlinks, linkcolor=blue, anchorcolor=blue, citecolor=blue, urlcolor=blue]{hyperref}
\usepackage{color}
\usepackage{bbold}
\usepackage{mathtools}
\usepackage[normalem]{ulem}
\usepackage{times}
\usepackage{xcolor}
\usepackage[capitalize]{cleveref}
\usepackage{siunitx}

\AtBeginDocument{\RenewCommandCopy\qty\SI}
\newcommand{\mum}{\,\si{\micro\meter}}

\newcommand{\rubidium}{^{87}\text{Rb}}

\def\Tr{\mbox{Tr}}

\makeatletter

\newcommand{\specificthanks}[1]{\@fnsymbol{#1}}

\def\convertto#1#2{\strip@pt\dimexpr #2*65536/\number\dimexpr 1#1}

\begin{document}

\title{Experimentally probing Landauer's principle in the quantum many-body regime}

\author{Stefan~Aimet}
\thanks{These authors contributed equally to this work: \\\href{mailto:stefan.aimet@gmail.com}{stefan.aimet@gmail.com}, \href{mailto:amintajik.physics@gmail.com}{amintajik.physics@gmail.com}}
\affiliation{Dahlem Centre for Complex Quantum Systems, Freie Universität Berlin, 14195 Berlin, Germany}
\author{Mohammadamin~Tajik}
\thanks{These authors contributed equally to this work: \\\href{mailto:stefan.aimet@gmail.com}{stefan.aimet@gmail.com}, \href{mailto:amintajik.physics@gmail.com}{amintajik.physics@gmail.com}}
\affiliation{Vienna Center for Quantum Science and Technology, Atominstitut, TU Wien, 1020 Vienna, Austria}
\author{Gabrielle~Tournaire}
\affiliation{Dahlem Centre for Complex Quantum Systems, Freie Universität Berlin, 14195 Berlin, Germany}
\affiliation{Department of Physics and Astronomy, and Stewart Blusson Quantum Matter Institute,
University of British Columbia, V6T1Z1 Vancouver, Canada}
\author{Philipp Schüttelkopf}
\affiliation{Vienna Center for Quantum Science and Technology, Atominstitut, TU Wien, 1020 Vienna, Austria}
\author{João~Sabino}
\affiliation{Vienna Center for Quantum Science and Technology, Atominstitut, TU Wien, 1020 Vienna, Austria}
\author{Spyros~Sotiriadis}
\affiliation{Institute of Theoretical and Computational Physics, University of Crete, 71003 Heraklion, Greece}
\affiliation{Dahlem Centre for Complex Quantum Systems, Freie Universität Berlin, 14195 Berlin, Germany}
\author{Giacomo~Guarnieri}
\affiliation{Dipartimento di Fisica,  Università di Pavia, 27100 Pavia, Italy}
\affiliation{Dahlem Centre for Complex Quantum Systems, Freie Universität Berlin, 14195 Berlin, Germany}
\author{Jörg~Schmiedmayer}
\affiliation{Vienna Center for Quantum Science and Technology, Atominstitut, TU Wien, 1020 Vienna, Austria}
\author{Jens~Eisert}
\affiliation{Dahlem Centre for Complex Quantum Systems, Freie Universität Berlin, 14195 Berlin, Germany}

\begin{abstract}
Landauer’s principle bridges information theory and thermodynamics by linking the entropy change of a system during a process to the average energy dissipated to its environment. Although typically discussed in the context of erasing a single bit of information, Landauer’s principle can be generalised to characterise irreversibility in out-of-equilibrium processes, such as those involving complex quantum many-body systems. 
Specifically, the relationship between the entropy change of the system and the energy dissipated to its environment can be 
decomposed into changes in quantum mutual information and a difference in relative entropies of the environment. 
Here we experimentally probe Landauer’s principle in the quantum many-body regime using a quantum field simulator of ultracold Bose gases. Employing a dynamical tomographic reconstruction scheme, we track the temporal evolution of the quantum field following a global mass quench from a massive to massless Klein-Gordon model and analyse the thermodynamic and information-theoretic contributions to a generalised entropy production for various system-environment partitions of the composite system. Our results verify the quantum field theoretical calculations, interpreted using a semi-classical quasiparticle picture.
Our work demonstrates the ability of ultracold atom-based quantum field simulators to experimentally investigate quantum thermodynamics.

\end{abstract}

\maketitle
Information theory and thermodynamics constitute foundational pillars of modern technology, underpinning our understanding of computers and heat engines, respectively. While these disciplines may seem distinct, they are intricately connected. This connection is encapsulated in Landauer's principle, first articulated in 1961~\cite{Landauer_original}. Landauer understood that erasing a bit of information in a computer is not for free but is instead accompanied by a minimal energy cost dissipated to the environment. Subsequently, various experimental studies~\cite{berut2012experimental, Hong2016, Yan2018} have confirmed this lower bound on energy dissipation for near-reversible bit erasure. Logical operations thus imply irreversibility, posing a fundamental theoretical limitation to the design of any small-scale energy-efficient information-processing technologies~\cite{lloyd2000ultimate}. 

However, extending beyond the case of bit erasure, recent influential work~\cite{reeb2014improved} has generalised the link between information theory and thermodynamics. Using a quantum statistical mechanics framework, their work reinterprets Landauer's principle as a means of relating the entropy change of a system to the energy dissipated to its environment in general out-of-equilibrium processes, not just erasure. This relationship can be quantified by a measure of process irreversibility~\cite{Landi2021}. Such a broader formulation of Landauer's principle not only deepens its physical significance but also makes this extension particularly relevant for quantum many-body systems, where contributions to irreversibility remain an area of active research, notably in phenomena such as equilibration and thermalisation~\cite{polkovnikov2011colloquium, eisert2015quantum, gogolin2016equilibration, abanin2019colloquium}. 

In this work, we employ Landauer's principle to experimentally characterise the irreversibility of an out-of-equilibrium process in the quantum many-body regime by tracking the time evolution of quantum information-theoretic measures. We present a crisp information-inspired interpretation of the correlations present in states of quantum many-body systems in terms of entropic expressions. 

Concretely, we consider a system-environment composite initially in the state $\varrho_{\rm SE}(0)$, evolving under a global unitary time evolution $U$ to the state $\varrho_{\rm SE}(t)=U\varrho_{\rm SE}(0)U^{\dagger}$. The reduced state of the system $\rm S$ (environment $\rm E$) is $\varrho_{\rm S(E)}(t)=\Tr_{E(S)}[\varrho_{\rm SE}(t)]$. 
First, let us assume that there are no initial correlations between the system and environment, and the environment is in a thermal state $\gamma_{\rm E}^{\beta_{\rm E}}=e^{-\beta_{\rm E}H_{\rm E}}/\Tr[e^{-\beta_{\rm E}H_{\rm E}}]$ at inverse temperature $\beta_{\rm E}$ with respect to its Hamiltonian $H_{\rm E}$. In this case, the entropy production $\Sigma(t)$, which is a measure of irreversibility~\cite{esposito2010entropy, reeb2014improved, Landi2021}, can be decomposed as
\begin{equation}\label{eq:Sigmat}
\Sigma(t):=I_{\rm SE}(t)+D(\varrho_{\rm E}(t)||\gamma_{\rm E}^{\beta_{\rm E}}) \;.
\end{equation}
Here, the quantum relative entropy
\begin{equation}\label{eq:RelEntropy}
D(\varrho_{\rm E}(t)||\gamma_{\rm E}^{\beta_{\rm E}})=\Tr[\varrho_{\rm E}(t)\log{\varrho_{\rm E}(t)}]-\Tr[\varrho_{\rm E}(t)\log{\gamma_{\rm E}^{\beta_{\rm E}}}] \; 
\end{equation}
quantifies the deviation of the environment from its initial thermal state. The quantum mutual information 
\begin{equation}\label{eq:mi}
I_{\rm SE}(t)=S(\varrho_{\mathrm{S}}(t))+S(\varrho_{\mathrm{E}}(t))-S(\varrho_{\mathrm{SE}}(t))
\end{equation}
measures the system-environment correlations, where 
\begin{equation}\label{eq:vNentropy}
S(\varrho)=-\Tr[\varrho\log{\varrho}]
\end{equation}
denotes the von Neumann entropy of a state $\varrho$.
\begin{figure*}
    \centering
    \includegraphics[width=0.95\textwidth]{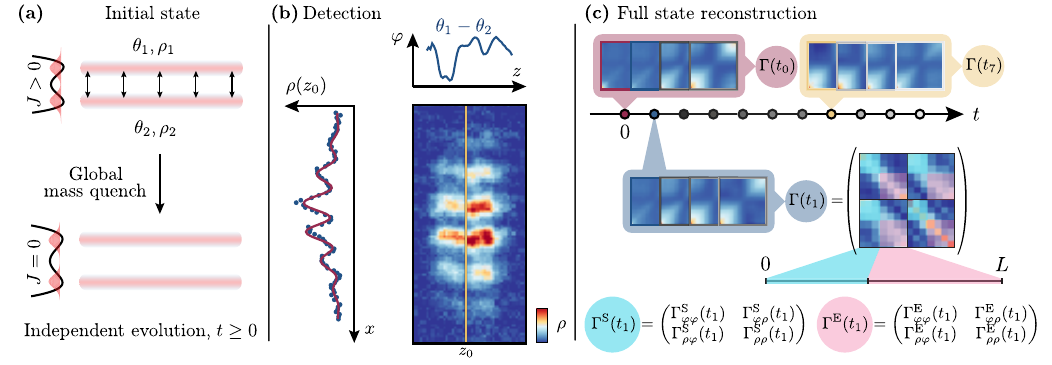}
    \caption{\textbf{Schematic of the experimental protocol.} \textbf{a,}  The experimental system consists of two tunnelling-coupled ultracold $\rubidium$ gases, with a single-particle tunnelling rate $J$, initially prepared in an initial state described by a global thermal state of the massive Klein-Gordon Hamiltonian. By ramping up a barrier between the condensates, a global mass quench is performed, and the condensates evolve independently under the post-quench massless Klein-Gordon Hamiltonian for $t\geq 0$. \textbf{b,} The atomic clouds are released, and they interfere as they expand. We obtain integrated 2D atomic density via absorption imaging, from which the relative phase profiles are obtained. As an  \textbf{c,} Using the measured phase-phase correlations, we dynamically reconstruct the full state. By successively shifting the observation window, we fit the covariance matrix $\Gamma(t)$ for 
    different times $t$. The covariance matrices of system $\rm S$ and environment $\rm E$ are defined accordingly.}
    \label{fig:setupandtomography}
\end{figure*}

Now we move on to a more general scenario where the system and the environment are initially correlated, and the state of the environment deviates from thermal equilibrium. We thereby introduce the \textit{generalised entropy production} $\Delta\Sigma :=\Sigma(t)-\Sigma(0)$ where the initial correlations and deviations from thermal equilibrium are accounted for by a resource cost $\Sigma(0)=I_{\rm SE}(0)+D(\varrho_{\rm E}(0)||\gamma_{\rm E}^{\beta_{\rm E}})$~\cite{mondal2023modified}, with $\beta_{\rm E}$ being an effective inverse temperature~\cite{kliesch2014locality, Lipka-Bartosik2023}. Landauer's principle can then be expressed~\cite{mondal2023modified} as an equality as 
\begin{equation}\label{eq:Landauer}
    \Delta\Sigma=\beta_{\rm E}\Delta E_{\rm E}+\Delta S=\Delta I+\Delta D \; ,
\end{equation}
relating the system's entropy 
change $\Delta S:=S(\varrho_{\rm S}(t))-S(\varrho_{\rm S}(0))$ to the energy dissipated to the environment $\Delta E_{\rm E}:=\Tr[(\varrho_{\rm E}(t)-\varrho_{\rm E}(0))H_{\rm E}]$. In this general formulation, depending on the initial state, $\Delta\Sigma$ is not necessarily non-negative~\cite{strasberg2019non,micadei2019reversing}.
Alternatively, $\Delta\Sigma$ also decomposes into the change of quantum mutual information, $\Delta I=I_{\rm SE}(t)-I_{\rm SE}(0)$, expressing how much the system-environment correlations change along the out-of-equilibrium process, as well as the term $\Delta D=D(\varrho_{\mathrm{E}}(t)||\gamma_{\mathrm{E}}^{\beta_{\mathrm{E}}})-D(\varrho_{\mathrm{E}}(0)||\gamma_{\mathrm{E}}^{\beta_{\mathrm{E}}})$ that quantifies how much both the initial and final state of the environment are different from the reference equilibrium thermal state. We provide a more thorough motivation for the generalised entropy production $\Delta\Sigma$ in the Methods section. 

Expressed in this form, Landauer's principle serves as a means of tracking changes in information-theoretic quantities, which compose generalised entropy production $\Delta\Sigma$ in two different ways via a single equation. Herein lies the power at the heart of thermodynamics: just as knowing the precise microscopic degrees of freedom, such as the position and momentum of each particle in a classical gas, is not useful for drawing meaningful conclusions about its dynamics, knowing the density matrix of a quantum many-body system is similarly not descriptive. In traditional thermodynamics, the relevant quantities that effectively characterise the dynamics are macroscopic variables like changes in pressure or volume. Similarly, for the out-of-equilibrium dynamics of a quantum many-body system, analysing the information-theoretic quantities in Landauer's principle allows for a simple characterisation of its irreversibility.

So far, there has been no systematic experimental pursuit of probing Landauer's principle in the quantum many-body regime. Here, we address this gap by measuring the time evolution of the different terms in Eq.~(\ref{eq:Landauer}) in a $(1+1)\mathrm{D}$ quantum field simulator of tunnelling-coupled ultracold Bose gases. This experimental platform has proven to be a suitable testbed for simulating one-dimensional quantum field theories. Previous studies using this setup have examined the area law of quantum mutual information in thermal equilibrium~\cite{tajik2023verification} 
and the propagation of second-~\cite{tajik2023experimental} or higher-order correlations~\cite{schweigler2021decay}. 

In this experiment, two parallel ultracold clouds of $\rubidium$ atoms (illustrated in Fig.~\hyperref[fig:setupandtomography]{\ref*{fig:setupandtomography}a}) are confined in highly anisotropic magnetic traps produced by an atom chip~\cite{Folman2000}. In the axial direction ($z$), the clouds are confined in a parabolic trap superimposed with an optical dipole potential to produce hard walls, while in the radial directions ($x,\, y$), by a double-well trap with an adjustable barrier in between, created by RF dressing~\cite{HLF06}. The single-particle tunnelling rate can be modified by the amplitude of a radio-frequency field created by two parallel wires on the atom chip.

The bosonic quantum field operator for each condensate can be written using the phase-density representation as $\psi_n(z) = \sqrt{\rho_n(z)}e^{i\theta_n(z)}$, where $\theta_n$ and $\rho_n$ denote the phase and density of the respective condensate, indexed by $n=1,2$. In the following, we focus on the operators $\varphi(z) = \theta_1(z) - \theta_2(z)$ and $\delta\rho(z) = [\rho_1(z) - \rho_2(z)]/2$ that represent the relative phase and relative density, respectively. These relative degrees of freedom, satisfy a similar commutation relation as the original phase and density operators of each condensate, given by $[\varphi(z), \delta\rho(z')] = i\delta(z-z')$. For strong tunnelling-coupling between the two clouds, the relative degrees of freedom are described by a massive \emph{Klein-Gordon} (KG) Hamiltonian,
\begin{equation} \label{eq:KG_maintext}
\begin{aligned}
	H_{\rm KG} &= \int_{0}^{L} dz \bigg[ g_{\rm 1D}\delta\rho^2(z) + \frac{\hbar^2 n_{\rm 1D}}{4m} (\partial_z\varphi(z))^2 \\
    &\hspace{4cm} + \hbar J n_{\rm 1D}\varphi^2(z) \bigg]
\end{aligned}
\end{equation}
which models low-energy phononic excitations, and is an approximation to an interacting sine-Gordon Hamiltonian~\cite{schweigler2017experimental} (see Supplementary Information for further details). In Eq.~(\ref{eq:KG_maintext}), $\hbar$ is the reduced Planck constant, $m$ is the atomic mass, $g_{\rm 1D}$ is the effective one-dimensional atomic interaction strength, $L=49\mum$ is the axial length of the condensates, $n_{\rm 1D} \approx 70\mum^{-1}$ is the average linear density and $J\approx 2\pi\times 0.8\, \si{Hz}$ is the tunnelling rate introduced before. Note that a mass term only appears due to the tunnelling-coupling between the pair of condensates. A single condensate can only simulate a \emph{massless} Tomonaga-Luttinger liquid model~\cite{luttinger1963exactly, Mora2003} ($H_\mathrm{KG}$ with $J=0$).

In our experiments, we prepare the Bose-Einstein condensates in a global thermal state of $H_{\rm KG}$ with finite tunnelling-coupling ($J>0$). Since we want to measure the out-of-equilibrium evolution of information-theoretic quantities, we drive the system out of equilibrium by rapidly quenching $J$ to zero (Fig.~\hyperref[fig:setupandtomography]{\ref*{fig:setupandtomography}a}). We do so by ramping up the barrier between the condensates within approximately $2\, \si{\milli \s}$. This change corresponds to a global mass quench of the Klein-Gordon Hamiltonian. The condensates then evolve independently under the post-quench massless Klein-Gordon Hamiltonian for times $t$,
up to $65\, \si{\milli \s}$.

\begin{figure}[t]
    \includegraphics[width=0.48\textwidth]{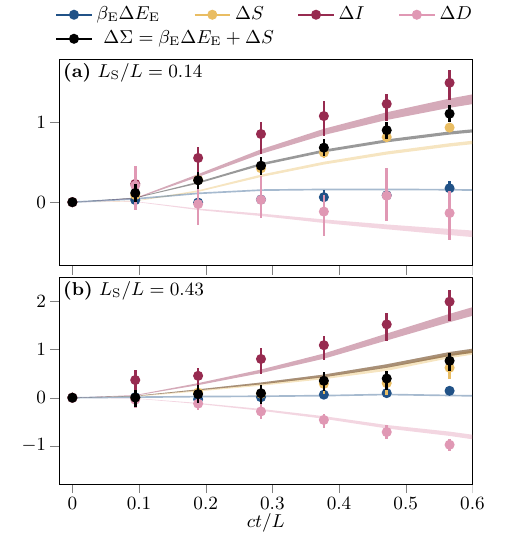}
    \caption{\textbf{Time evolution of different quantities in Landauer's principle.} The quantities of Landauer's principle are shown as a function of time for subregion size ratios \textbf{a,} $L_{\rm S}/L=0.14$ and \textbf{b,} $L_{\rm S}/L=0.43$. Experimental data is represented by circles with
    error bars marking the 68\% confidence intervals obtained via bootstrapping with $999$ samples. The shaded areas show the 68\% confidence interval for the theoretical predictions, considering the uncertainty in the estimated temperature and tunnelling rate obtained via bootstrapping with $999$ samples. The experimental data is in agreement with the quantum field theory simulation results using Neumann boundary conditions and considering the finite imaging resolution.
    }
    \label{fig:time_evolution_main}
\end{figure}

\begin{figure}[t]
    \centering
    \includegraphics[width=0.48\textwidth]{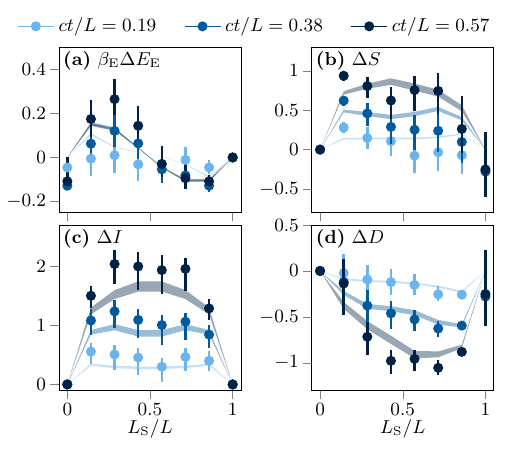}
    \caption{\textbf{Scaling of different quantities in Landauer's principle.} The quantities of Landauer's principle are shown as a function of subregion size for various times $ct/L=0.19, 0.38, 0.57$ (see Fig.~\ref{fig:time_evolution_main} for more details on the error bars and shaded areas). The experimental data is in agreement with the quantum field theory simulation results using Neumann boundary conditions.}
    \label{fig:scaling_main}
\end{figure} 

For every time step $t$, we turn off all the traps and let the atoms fall freely for $15.6\, \si{\milli \s}$. The clouds will expand and interfere, resulting in absorption pictures similar to the one shown in Fig.~\hyperref[fig:setupandtomography]{\ref*{fig:setupandtomography}b}, allowing us to measure the spatially resolved relative phase $\varphi(z)$ between them. Since the detection process is destructive, the measurements are repeated to gather statistics. In our current experimental setup, relative density fluctuations $\delta\rho$ are not directly measurable, which prompted the development of a \emph{dynamical tomographic reconstruction technique}~\cite{gluza2020quantum, tajik2023verification} to access all of the elements of the covariance matrix 
\begin{equation}
\Gamma(t)=\begin{pmatrix}
   \Gamma_{\varphi\varphi}(t) & \Gamma_{\varphi\rho}(t)\\
   \Gamma_{\rho\varphi}(t)& \Gamma_{\rho\rho}(t)\\
\end{pmatrix}\; .
\end{equation}
Here, the elements are defined as $[\Gamma_{\varphi\varphi}(t)]_{m,n}=\langle\varphi(z_m, t)\varphi(z_n, t)\rangle$, $[\Gamma_{\rho\rho}(t)]_{m,n}=\langle\delta\rho(z_m,t)\delta\rho(z_n,t)\rangle$, and $[\Gamma_{\varphi\rho}(t)]_{m,n}=[\Gamma_{\rho\varphi}(t)]^T_{m,n}=\langle\frac{1}{2}\{\varphi(z_m, t),\delta\rho(z_n, t)\}\rangle$, all on a discrete grid with $N$ pixels ($m,n\in\{1,\ldots,N\}$), which are determined by resolution constraints of our imaging system, thus introducing an ultraviolet (UV) cutoff.

The phase–phase correlations for different evolution times after the quench can be measured directly from the extracted relative phase profiles. Assuming that the short-time dynamics after the quench are governed by the massless Klein-Gordon Hamiltonian, as time progresses, the initial eigenmodes of the relative density transform into the phase quadrature, and the phase quadrature transforms into the relative density. This transformation allows us to extract information about these eigenmodes by fitting the initial second-order correlation functions of phase–density, and density–density with the observed evolution of the phase–phase correlations in momentum space. 

The schematics in Fig.~\hyperref[fig:setupandtomography]{\ref*{fig:setupandtomography}c} illustrate our dynamical tomographic reconstruction scheme. We follow the technique in Ref.~\cite{tajik2023verification} for various input intervals with varying starting points. By scanning the starting points of these input intervals throughout the trapping times, we can reconstruct the full covariance matrix at every time $t$, as depicted in Fig.~\ref{fig:setupandtomography}. The interval length ($32.5\, \si{\milli \s}$ close to $L/c \approx 27\, \si{\milli \s}$) is selected to be sufficiently long for the slowest eigenmode to acquire enough dynamical phase for a stable reconstruction, yet short enough to prevent mode interactions from affecting the reconstruction. A detailed overview of this reconstruction process is provided in the Methods section.

The quadratic form of the pre- and post-quench Hamiltonians allows us to work within the framework of Gaussian quantum information theory~\cite{Weedbrook2012, Continuous}. In this framework, the covariance matrix $\Gamma$ captures all the accessible information about the state of the composite system during the dynamics, from which we can extract all the information-theoretic quantities of interest in Eq.~(\ref{eq:Landauer}). As has been demonstrated in various experiments~\cite{langen2013local, langen2015experimental, Yang2017, schweigler2017experimental, rauer2018recurrences}, the quadratic approximation of the Hamiltonian accurately captures the dynamics for the considered time scales. Even if the true dynamics deviate from the Gaussian regime and violate the massless Klein-Gordon theory, a Gaussian extremality argument presented in the Supplementary Information justifies the Gaussian tomography scheme and provides bounds on the information-theoretic quantities. Thus, based on our technique of dynamical tomographic reconstruction of covariance matrices, Landauer's principle can also be meaningfully experimentally investigated for interacting models.

Having experimentally reconstructed the post-quench time evolution of the covariance matrices, we partition the one-dimensional field of length $L$ into two distinct subregions and split the covariance matrix accordingly (Fig.~\hyperref[fig:setupandtomography]{\ref*{fig:setupandtomography}c}). Since the quantum field is isolated from its surroundings, one subregion serves as the system $\mathrm{S}$ with length $L_{\mathrm{S}}$, while the other subregion functions as the environment $\mathrm{E}$ with length $L_{\mathrm{E}} = L - L_{\mathrm{S}}$. We probe Landauer's principle (Eq.~(\ref{eq:Landauer})) for various system-environment bipartitions by characterising the generalised entropy production $\Delta\Sigma$. To do so, we compute the individual contributions $\beta_{\rm E}\Delta E_{\rm E}, \Delta S, \Delta I$ and $\Delta D$.

The results are presented in Fig.~\ref{fig:time_evolution_main} as a function of time for different subregion size ratios and in Fig.~\ref{fig:scaling_main} as a function of subregion size for different times. The time scales are shown in units of $ct/L$, where $c=\sqrt{g_{\rm 1D}n_{\rm 1D}/m}\approx 1.8\, \si{\micro \meter} / \si{\milli \second} $ is the speed of sound. Overall, a very good fit of the data (circles) is obtained compared to the theoretical calculations (shaded areas), which use the lowest $N=7$ modes and consider the imaging resolution of the experiment. The low-lying modes already capture the dynamics of the continuum theory very well. The error bars representing the $68\%$ confidence intervals are obtained via bootstrapping~\cite{efron1986bootstrap} with 999 samples, considering the uncertainty in the tunnelling rate and the estimated initial global temperature. The effective inverse temperature of the environment $\beta_{\rm E}$ with respect to the post-quench massless Klein-Gordon Hamiltonian is computed using quantum field-theoretic simulations by constructing a global thermal state of the pre-quench massive Klein-Gordon Hamiltonian with the estimated initial global temperature from the experimental data. 

First, we examine the decomposition of the generalised entropy production $\Delta\Sigma$ into $\beta_{\rm E}\Delta E_{\rm E}$ and $\Delta S$. The term associated with the energy dissipated to the environment, $\beta_{\rm E}\Delta E_{\rm E}$, reveals that a small amount of energy flows into (or out of) the environment for small (or large) systems around $ct/L_{\rm S}=1$. However, this term contributes only minimally to the irreversible post-quench dynamics of the quantum field. In contrast, the entropy change of the system, $\Delta S$, dominates the dynamics and shows a clear linear increase up to $ct/L_{\rm S} = 1$, followed by more pronounced growth. Notably, in Fig.~\ref{fig:time_evolution_main}, we present only $\Delta\Sigma$ defined by this first decomposition. However, we also find good agreement with the second decompositon (see Supplementary Information), thus providing evidence that the assumptions underlying Landauer's principle are well satisfied in our experiment.

In comparison to the first, the second decomposition involving $\Delta I$ and $\Delta D$ sheds a slightly different focus on the out-of-equilibrium dynamics. The largest contribution to $\Delta\Sigma$, the change in quantum mutual information, $\Delta I$, exhibits similar behaviour to $\Delta S$ but with a higher magnitude. This increased magnitude is accounted for by the additional entropy change in the environment. However, $\Delta I$ competes with the term proportional to the change of the environment, $\Delta D$, which accounts for both the entropic and energetic changes to the environment. This makes $\Delta D$ a particularly interesting quantity to characterise the out-of-equilibrium dynamics of the environment experimentally. In our case, $\Delta D$ decreases, mirroring otherwise the behaviour of $\Delta S$, due to the small energetic contribution. 

\begin{figure}[t]
    \centering
    \includegraphics[width=0.49\textwidth]{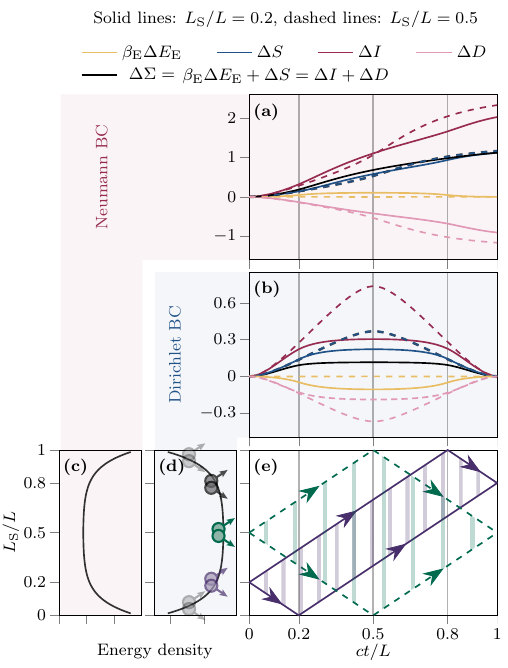}
    \caption{\textbf{Quasiparticle picture interpretation based on quantum field theory simulations.} The time evolution of the quantities in Landauer's principle is shown for \textbf{a,} Neumann boundary conditions (BC) (as relevant in the experiment), and \textbf{b,} Dirichlet boundary conditions, using quantum field-theoretic simulations. We interpret the post-quench dynamics of the global mass quench using a semi-classical quasiparticle picture, as discussed in the main text. \textbf{c},-\textbf{d,} The energetic dynamics can be explained by the differing energy density of the initial state at the edges compared to the bulk. \textbf{e,} The linear increase in correlations for $ct/L_{\rm S} < 1$, followed by a plateau, and a linear decrease for $ct/L_{\rm E} > 1$, can be explained by the linear effective light cone originating at the system-environment boundary. The effect of the zero-mode, present in Neumann boundary conditions, is not captured by the quasiparticle picture for $ct/L_{\rm S}>1$.}
    \label{fig:DBC_main}
\end{figure}

To interpret the results, we need to consider the effects of Neumann and Dirichlet boundary conditions on the quantum field, as shown with quantum field-theoretic simulations in Fig.~\ref{fig:DBC_main}. These boundary conditions may be contrasted with the curved
backgrounds giving rise to effective boundary conditions discussed in Refs.~\cite{tajik2023experimental, PhysRevA.78.033608}. The experimental system exhibits Neumann boundary conditions ($\partial_z\varphi(z)|_{z=0, L}=0$) due to the vanishing particle current at the edges. Plots contrasting the scaling with subregion size are shown in the Supplementary Information.

However, we first consider simulations for Dirichlet boundary conditions ($\varphi(z)|_{z=0, L}=0$), which are simpler to understand. Here, the well-established semi-classical quasiparticle picture~\cite{calabrese2005evolution} offers a clear and intuitive model for the post-quench dynamics of global mass quenches~\cite{calabrese2016quantum, di2020entanglement}, depicting the propagation of short-range initial correlations via ballistically moving quasiparticles. Following the homogeneous global quench, the initial massive Klein-Gordon thermal state becomes a non-equilibrium state with excess energy relative to the post-quench massless Klein-Gordon Hamiltonian, which governs the time evolution. This initial state acts as a source of quasiparticle pairs emitted globally from every point across the length of the composite system. Each pair in the bulk consists of two quasiparticles correlated with each other and moving in opposite directions at the same speed. For our set of parameters, due to the finite initial correlation length, there exists a small localised region where quasiparticles are correlated.

In this quasiparticle picture, the change of quantum mutual information $\Delta I$ between the system and environment is proportional to the number of pairs of quasiparticles that are shared between the two different subregions. The spread of correlations of short-range interacting models is described by a linear effective light cone originating at the system-environment boundary. The increase in correlations is thus proportional to the section of the light cone, or in other words, it is proportional to the distance between the quasiparticles of the pair emitted at the system-environment boundary (Fig.~\hyperref[fig:DBC_main]{\ref*{fig:DBC_main}e}). If the system is smaller than the environment, once the first quasiparticle of this boundary pair reaches the edge of the system at $ct/L_{\rm S}=1$ and undergoes reflection, the section of the system-environment boundary pair light cone remains constant. The behaviour of $\Delta I$ transitions to a plateau value. From a system perspective, the composite system appears to have locally equilibrated to a generalised Gibbs ensemble steady state~\cite{sotiriadis2014validity, langen2015experimental, calabrese2018entanglement}. For finite-size composite systems, the plateau eventually ends again at $ct/L_{\rm E}=1$ when the other quasiparticle of the boundary pair reaches the edge. The distance between the two quasiparticles decreases until they meet again at $ct/L=1$, explaining the phenomenon of recurrences~\cite{rauer2018recurrences, modak2020entanglement} (with respect to the correlation functions). 

On the other hand, $\Delta I$ behaves differently for Neumann boundary conditions. Neumann boundary conditions introduce a zero mode, whose variance does not evolve harmonically like the other momentum modes, but quadratically (phase diffusion~\cite{Lewenstein1996, Jo2007}). The global fluctuations of the zero mode contribute to entropic quantities~\cite{michel2016entanglement} and as they increase with time after a quench they eventually dominate over the contribution of other modes. Since the quasiparticle picture cannot capture this zero-mode feature, its predictions for $ct/L_{\rm S}>1$, when a plateau value should be reached, need to be modified under the present Neumann boundary conditions. The above features of the dynamics also apply to the change of system entropy $\Delta S$. In addition, the zero mode restricts the validity of the tomography scheme to $ct/L<1$. 

The energetic contribution, $\beta_{\rm E}\Delta E_{\rm E}$, can also be explained using the quasiparticle picture. Initially, $\beta_{\rm E}\Delta E_{\rm E}$ remains constant because quasiparticles traveling between the system and the environment carry equal energy, resulting in zero net energy flux. However, a finite-size energy flow becomes apparent when quasiparticles from the edge of the system cross the system-environment boundary. The translational invariance of the homogeneous quench is broken due to the higher energy density at the edges of the composite system for Neumann boundary conditions (the reverse is true for Dirichlet boundary conditions, where the energy density is lower at the edges). This edge region has a size of the order of the initial correlation length. The resulting behaviour of the change of the environment $\Delta D$ reflects both the entropic and energetic fluxes. 

In this study, we have experimentally probed Landauer's principle in the quantum many-body regime following a global mass quench in an ultracold atom-based quantum field simulator. By reconstructing the dynamics of the state of the composite system, we examined the information-theoretic quantities related by Landauer's principle, which we interpret using a semi-classical quasiparticle picture. Our approach underscores the general utility of Landauer's principle for characterising the irreversibility of out-of-equilibrium dynamics in quantum many-body systems. The way we express correlations in terms of entropic quantities with an 
information-theoretic meaning can be viewed as a vehicle to capture many-body correlations in 
non-equilibrium. The Gaussian extremality argument may offer a pathway to extend our methodology to capture features of out-of-equilibrium processes in interacting models, despite the challenges posed by non-Gaussian effects. Looking forward, progress has already been made towards investigating a local quench involving two Bose-Einstein condensates of different temperatures being joined together~\cite{ventura2024quantum}. This protocol represents a crucial primitive towards developing a quantum field thermal machine~\cite{QFM}, potentially allowing for Landauer erasure as a mechanism to reduce the entropy of a subregion in the quantum many-body regime, hence functioning as an effective cooling mechanism. Our current work demonstrates the potential of ultracold one-dimensional gases as testbeds for quantum thermodynamics in the many-body regime, where complexity, quantum effects, and finite-size effects play a crucial role.

\section*{Data availability}
The data required to extract and calculate the presented results in Figs. 1-3 are available from the corresponding authors upon reasonable request.

\acknowledgements
We thank Bernhard Rauer for help with the initial set of measurements. In addition, we are grateful to Marcus Huber, Patrick Emonts, Ivan Kukuljan, Mauro Paternostro, Philippe Faist, Florent Goulette, Marek Gluza, Igor Mazets, Silke Weinfurtner, and Maciej Jarema for helpful discussions and comments. This work has been supported by the DFG Research Unit FOR 2724 on `Thermal machines in the quantum world', the FQXi, the Quantum Flagship (`Millenion' and `PasQuans2'), the Einstein Research Unit, the BMBF, Berlin Quantum, and the ERC-AdGs `Emergence in Quantum Physics' and 
`Delineating the boundary
between the computational
power of quantum and
classical devices'.
S.S.\ acknowledges support by
the European Union's Horizon 2020 research and innovation program under the Marie Skłodowska-Curie Grant Agreement No.\  101030988. 

\section*{Methods}

\subsection*{Landauer's principle}
Consider a system $\mathrm{S}$ and an environment $\mathrm{E}$ with a bipartite Hilbert space $\mathcal{H}=\mathcal{H}_{\mathrm{S}}\otimes\mathcal{H}_{\mathrm{E}}$. For an initial product state $\varrho_{\mathrm{SE}}=\varrho_{\mathrm{S}}\otimes \gamma_{\mathrm{E}}^{\beta_{\mathrm{E}}}$, where the environment is in a thermal state $\gamma_{\mathrm{E}}^{\beta_{\rm E}}=e^{-\beta_{\rm E} H_{\rm E}}/\Tr{[e^{-\beta_{\mathrm{E}} H_{\mathrm{E}}}]}$, the entropy production under global unitary dynamics is given by~\cite{Landi2021}
\begin{equation}
\Sigma(t)=I_{\rm SE}(t)+D(\varrho_{\mathrm{E}}(t)||\gamma_{\mathrm{E}})=D(\varrho_{\rm SE}(t)||\varrho_{\mathrm{S}}(t)\otimes \gamma_{\mathrm{E}}^{\beta_{\rm E}})\; .
\end{equation}
Since it is expressed as a quantum relative entropy, $\Sigma$ is non-negative due to Klein's inequality~\cite{klein1931quantenmechanischen}, which is a manifestation of the second law of thermodynamics. More generally, as in our work, we start with a \textit{resourceful} initial state that includes both initial correlations $I_{\rm SE}(0)$ and athermality of the environment $D(\varrho_{\mathrm{E}}(0)||\gamma_{\mathrm{E}}^{\beta_{\rm E}})$. While defining temperature for a non-equilibrium state is ambiguous, we determine the effective inverse temperature of the environment $\beta_{\rm E}$ by demanding that the environment's energy is equal to that 
of a Gibbs state, such that $\Tr[\varrho_{\rm E}H_{\rm E}]\equiv\Tr[\gamma_{\mathrm{E}}^{\beta_{\rm E}}H_{\rm E}]$~\cite{strasberg2021first}. This choice also minimises the athermality term. Thus, we define the resource cost of the initial state~\cite{mondal2023modified}
\begin{equation} 
\Sigma(0)=I_{\rm SE}(0)+D(\varrho_{\mathrm{E}}(0)||\gamma_{\mathrm{E}}^{\beta_{\rm E}})=D(\varrho_{\rm SE}(0)||\varrho_{\mathrm{S}}(0)\otimes\gamma_{\rm E}^{\beta_{\rm E}}).
\end{equation}

Although there is no unique consensus on entropy production for initial non-equilibrium environment states~\cite{santos2019role} and initial correlations~\cite{bera2017generalized, jiang2018improved}, we define Landauer's principle Eq.~(\ref{eq:Landauer}) in terms of generalised entropy production $\Delta \Sigma:=\Sigma(t)-\Sigma(0)$ as
\begin{equation}
    \Delta \Sigma :=\beta_{\mathrm{E}}\Delta E_{\rm E}+\Delta S=\Delta I+\Delta D\; .
\end{equation}
This can be derived similarly to Ref.~\cite{reeb2014improved}. One caveat is that $\Delta\Sigma$ may not necessarily be non-negative, which is a necessary criterion for a strictly valid definition of entropy production. However, this makes sense as $\Delta\Sigma$ may be considered a finite-difference version of the entropy production rate $\Delta\Sigma=\Sigma(t)-\Sigma(0)=\int_0^{t} dt' \frac{d\Sigma(t')}{dt'}$, which can also be negative~\cite{strasberg2019non, micadei2019reversing}. Thus, $\Delta \Sigma$ intuitively measures the "extra" irreversibility associated with the protocol, excluding the initial resources.

\subsection*{Experimental implementation}
The experimental setup consists of a pair of tunnelling-coupled one-dimensional $\rubidium$ quasicondensates, following a protocol implemented previously~\cite{tajik2023verification}. The magnetic fields required to trap the gas in a double-well potential are generated by an atom chip~\cite{Folman2000}, achieving transverse trapping frequencies of $\omega_{\perp}/2\pi = 1.4 \text{kHz}$. After preparing a thermal equilibrium state at a temperature of $49 \text{nK}$, all traps are turned off, and absorption imaging of the atoms is performed during a free-fall time of $15.6 \text{ms}$ to measure the two-dimensional atomic density distributions. This procedure enables us to detect the relative phase between the condensates across the length of the quasicondensate using matter interferometry (Fig.~~\hyperref[fig:setupandtomography]{\ref*{fig:setupandtomography}b}). Due to the destructive nature of the detection process, many repetitions of the experiment are necessary to accumulate numerous phase correlation images throughout the time evolution, thereby also ensuring statistically accurate expectation values.

\subsection*{Quantum field simulation}
As discussed in previous works~\cite{schweigler2017experimental, tajik2023verification}, two tunnelling-coupled quasicondensates can be used as a quantum field simulator of the one-dimensional sine-Gordon model. The relative degrees of freedom of two conjugate variables, the relative phase ($\varphi$) and relative density flucutations ($\delta\rho$), in the regime of low-energy collective excitations can be described by the \emph{sine-Gordon} (sG) Hamiltonian:
\begin{equation}\label{eq:H_sG}
\begin{aligned}
    H_{\rm sG} = \int_{0}^{L} \!\!dz \bigg[ g_{\rm 1D}\delta\rho(z)^2 
    & - 2\hbar Jn_{\rm 1D} \cos{(\varphi(z)^2)} \\
    & + \frac{\hbar^2n_{\rm 1D}}{4m}(\partial_z\varphi(z))^2 \bigg] \;.
\end{aligned}
\end{equation}
In our experiment, the length of the composite system is set to $L =49\mum$, with a mean density of $n_{\rm 1D} = 70\mum^{-1}$. The coupling strength between the particles is chosen as $g_{\rm 1D} = 8.594\times 10^{-39}\text{kg}\cdot \text{m}^3\cdot \text{s}^{-2}$, and we have an atomic mass of $m = 1.433\times 10^{-25}\text{kg}$. The initial single-particle 
tunnelling rate is $J= 2\pi\times 0.76\text{Hz}$.
Under these parameters, the experimental system is sufficiently cold and strongly coupled, ensuring that the phase coherence length,
\begin{equation}
    \lambda_{\rm T}=\frac{2\hbar^2 n_{\rm 1D}}{m k_B T}\;,
\end{equation}
exceeds the healing length of the relative phase,
\begin{equation}
    l_{\rm C}=\sqrt{\frac{\hbar}{4mJ}}\;.
\end{equation}
Hence, the cosine term in Eq.~(\ref{eq:H_sG}) can be expanded to second order, reducing the sine-Gordon Hamiltonian to the Klein-Gordon Hamiltonian
\begin{equation}
\begin{aligned}\label{eq:H_KG_SI}
    H_{\rm KG} = \int_{0}^{L} \!\!dz \bigg[ g_{\rm 1D}\delta\rho(z)^2 
    & +\hbar Jn_{\rm 1D} \varphi(z)^2) \\
    & + \frac{\hbar^2n_{\rm 1D}}{4m}(\partial_z\varphi(z))^2 \bigg] \;.
\end{aligned}
\end{equation}
The post-quench evolution is then described by the massless Klein-Gordon Hamiltonian with $J=0$.

\subsection*{Dynamical tomographic reconstruction of covariance matrices}
The tomography scheme for reconstructing the initial thermal covariance matrix is detailed in Refs.~\cite{gluza2020quantum, tajik2023verification}. Here, we extend this method to dynamically reconstruct the state over time. We provide a brief overview below.
Given Neumann boundary conditions, $\partial_z\varphi(z)|_{z=0,L}=0$, in a box-like potential with length $L$ and average linear atomic density $n_\mathrm{1D}$, the relative phase $\varphi(z)$ and relative density fluctuations $\delta\rho(z)$ can be expanded in Fourier space in terms of cosine eigenfunctions
\begin{equation}\label{eq:eigenfunctions}
\begin{aligned}
f_k^{\varphi}(z) = \begin{cases}
2\left(\frac{1}{k\hbar\pi}\sqrt{\frac{mg_{\rm 1D}}{n_{\rm 1D}}}\right)^{1/2}\cos{\left(k\frac{\pi}{L}z\right)} & \text{if } k > 0 ,\\
1 & \text{if } k = 0,
\end{cases}
\\
f_k^{\rho}(z) = \begin{cases}
\frac{1}{L}\left(k\hbar\pi\sqrt{\frac{n_{\rm 1D}}{mg_{\rm 1D}}}\right)^{1/2}\cos{\left(k\frac{\pi}{L}z\right)} & \text{if } k > 0 ,\\
\frac{1}{L} & \text{if } k = 0.
\end{cases}
\end{aligned}
\end{equation}

The post-quench massless Klein-Gordon Hamiltonian (Eq.~\ref{eq:H_KG_SI}) is diagonalised as 
(indices omitted for brevity from now on)
\begin{equation}
    H=\frac{\hbar u}{2}\delta\rho_0^2+\sum_{k=1}^{\infty}\frac{\hbar\omega_k}{2}[\delta\rho_k^2+\varphi_k^2]
\end{equation}
with $u:=\frac{2g_{\rm 1D}}{\hbar L}$ and 
\begin{equation}
\omega_k:=\sqrt{\frac{g_{\rm 1D}n_{\rm 1D}}{m}}k\frac{\pi}{L}\; .
\end{equation}
The interaction strength $g_{\rm 1D}$ is given by 
\begin{equation}
g_{\rm 1D}:=\hbar\omega_{\perp}a_s\frac{2+3a_s n_{\rm 1D}}{(1+2a_s n_{\rm 1D})}\; ,
\end{equation}
where $a_s=5.2 \text{nm}$ is the three-dimensional scattering length.
Consequently, the time evolution of modes ($k>0$) in Fourier space is
\begin{equation}
\begin{aligned}
\varphi_k(t)=\varphi_k(0)\cos{(\omega_k t)}+\delta\rho_k(0)\sin{(\omega_k t)},\\
\delta\rho_k(t)=\delta\rho_k(0)\cos{(\omega_k t)}-\varphi_k(0)\sin{(\omega_k t)},
\end{aligned}
\end{equation}
with non-harmonic time evolution of the zero mode
\begin{equation}
\begin{aligned}
    \varphi_0(t)= ut\delta\rho_0(t)+\varphi_0(0),\\ \delta\rho_0(t)=\delta\rho_0(0)=\rm const \, .
\end{aligned}
\end{equation}
As mentioned in the main text and shown in Fig.~\ref{fig:setupandtomography}, the covariance matrices can be dynamically tomographically reconstructed by measuring the relative phase profiles using matter interferometry at various times. Subsequently, the relative phase correlations are computed as
$\Phi^2_{m,n}(t)=\langle(\varphi(z_m,t)-\varphi(z_0, t))(\varphi(z_n,t)-\varphi(z_0, t))\rangle$ with $z_0$ as a reference position. 
We then expand in Fourier space, see Eq.~(\ref{eq:eigenfunctions}), and expand the time evolved operators to express the measured phase correlation function in terms of the covariance matrix elements prior to the time evolution
\begin{equation}
\begin{aligned}
\Phi^2_{m,n}(t)=\langle(\varphi(z_m,t)-\varphi(z_0, t))(\varphi(z_n,t)-\varphi(z_0, t))\rangle\\
=\sum_{k,l=1}^{N-1}f_{k,l}^{m,n}\langle\varphi_k(t)\varphi_l(t)\rangle
\end{aligned}
\end{equation}
where
\begin{equation}
    f_{k,l}^{m,n}=(f_k^{\varphi}(z_m)-f_k^{\varphi}(z_0))(f_l^{\varphi}(z_n)-f_l^{\varphi}(z_0))
\end{equation}
Inserting the time evolution equations, we see that the covariance matrix $\Gamma(t)$ can be obtained by fitting the elements $\Gamma_{\varphi\varphi; k,l}(t)$, $\Gamma_{\rho\rho; k,l}(t)$, and $\Gamma_{\varphi\rho; k,l}(t)$ using the phase-space correlations
\begin{equation}
	\begin{aligned}
		\Phi^2_{m,n}(t+\Delta t)
		=\sum_{k,l=1}^{N-1}f_{k,l}^{m,n}\cos{(\omega_k \Delta t)}\cos{(\omega_l\Delta t)} \Gamma_{\varphi\varphi; k,l}(t)\\+\sum_{k,l=1}^{N-1}f_{k,l}^{m,n}\sin{(\omega_k \Delta t)}\sin{(\omega_l \Delta t)}\Gamma_{\rho\rho; k,l}(t)\\+\sum_{k,l=1}^{N-1}(f_{k,l}^{m,n}+f_{l,k}^{m,n})\cos{(\omega_k \Delta t)}\sin{(\omega_l \Delta t)}\Gamma_{\varphi\rho; k,l}(t)\; .
\end{aligned}
\end{equation}
By gathering statistics, we employ the same method as in Ref.~\cite{tajik2023verification}, but simply shift the window of observation.

We obtain the covariance matrix in momentum space for the first $N$ modes that give physical (positive) occupation numbers. The imaging system is additionally accounted for by a Gaussian point spread function with standard deviation of $3\mu m$ to estimate the initial state. In this way, we can also confirm the global initial thermal state and extract a pre-quench temperature. We use discrete cosine eigenfunctions to convert the momentum space correlators to real-space correlators, from which we compute the quantities of interest.

\subsection*{Probing Landauer's principle from covariance matrices}
In this section, we delineate how the quantities in Landauer's principle, as expressed in Eq.~(\ref{eq:Landauer}), are computed using the full covariance matrix $\Gamma$ in real space, defined on a grid of $N$ pixels.
First, the von Neumann entropy $S(\Gamma)$ is computed using the symplectic eigenvalues $\lambda_n$ of the covariance matrix $\Gamma$~\cite{Weedbrook2012}. These eigenvalues correspond to the eigenvalues of $i\Omega \Gamma$, where 
\begin{equation} 
\Omega:=\begin{pmatrix}
    0&\mathbb{1}_{N}\\
    -\mathbb{1}_{N}&0
\end{pmatrix} 
\end{equation}
is the 
standard kernel of the symplectic form capturing the canonical commutation relations. The expression for 
the von Neumann entropy 
$S(\Gamma)$ of a Gaussian state is given by
\begin{equation} S(\Gamma)=\sum_{n=1}^{N}\left[(\lambda_n+\frac{1}{2})\ln{(\lambda_n+\frac{1}{2})}-(\lambda_n-\frac{1}{2})\ln{(\lambda_n-\frac{1}{2})}\right].
\end{equation}

Given two full covariance matrices corresponding to different times and partitioning it into system and environment subspace covariance matrices (Fig.~\hyperref[fig:setupandtomography]{\ref*{fig:setupandtomography}c}), this allows us to compute the entropy change of the system $\Delta S$ and the mutual information $\Delta I$.

Next, we illustrate how to calculate the energy of the environment with respect to the post-quench Hamiltonian. To obtain the correct environment Hamiltonian, we need to proceed with care. The field corresponding to the environment is discretely defined with $N_{\rm E}=N (L_{\rm E}/L)$ 
pixels. A naive approach would discretise the continuous Hamiltonian defined over the length of the environment, but this nearest-neighbour phase coupling Hamiltonian would yield a non-linear dispersion relation, inadequately describing the physics of the environment for the continuum model used in the experiment. Instead, we start from the environment Hamiltonian of the continuous theory, diagonalise it to obtain a momentum decomposition of infinitely many modes, and then truncate the Hamiltonian~\cite{ott2024hamiltonian} of the environment to a form with a 
linear dispersion relation, retaining only $N_{\rm E}$ modes
\begin{equation}
    H_{\rm E}=\frac{\hbar u_{\rm E}}{2}(\delta\rho^{\rm E}_0)^2+\sum_{k=1}^{N_{\rm E}-1}\frac{\hbar\omega_k^{\rm E}}{2}[(\delta\rho_k^{\rm E})^2+(\varphi_k^{\rm E})^2]
\end{equation}
with $u_{\rm E}:=\frac{2g_{\rm 1D}}{\hbar L_{\rm E}}$ and $\omega_k^{\rm E}:=\sqrt{\frac{g_{\rm 1D}n_{\rm 1D}}{m}}k\frac{\pi}{L_{\rm E}}$. Converting back to real-space operators, we derive an effective resolution scale-dependent Hamiltonian
\begin{equation}
\begin{aligned}
    H_{\rm E}=g_{\rm 1D}\Delta z\sum_{n=N_{\rm S}}^{N-1}\delta\rho(z_n)^2\\+\frac{\hbar^2 n_{\rm 1D}}{4m\Delta z} \sum_{n,m=N_{\rm S}}^{N-1}\varphi(z_n)\varphi(z_m)\frac{2\pi^2}{N_{\rm E}^3}\kappa_{n,m}^{\rm E}\; ,
\end{aligned}
\end{equation}
where 
\begin{eqnarray}
\kappa_{n,m}^{\rm E}&:=&\sum_{k=1}^{N_{\rm E}-1} \Bigg\{ k^2\cos{(k\frac{\pi}{N_{\rm E}}(n-N_{\rm S}+\frac{1}{2}))} \\
\nonumber
&\times & \cos{(k\frac{\pi}{N_{\rm E}}(m-N_{\rm S}+\frac{1}{2}))} \Bigg\}
\end{eqnarray}
encodes the long-range coupling of the effective Hamiltonian. Here, $\Delta z:=L/N$ is the pixel size, $N_{\rm S(E)}$ is the number of pixels of the system (environment).

The energy of the environment can then be calculated~\cite{Weedbrook2012} as $\Tr[\varrho_{\rm E} H_{\rm E}]=\Tr[\Gamma^{\rm E}\tilde{H}_{\rm E}]$, where $\tilde{H}_{\rm E}$ is the matrix expression related to $H_{\rm E}=X^T\tilde{H}_{\rm E}X$, with $X=(\varphi(z_{N_S}), \dots,  \varphi(z_{N-1}), \delta\rho(z_{N_S}), \dots, \delta\rho(z_{N-1}))^T$. Thus, to calculate the energetic contribution $\beta_{\rm E}\Delta E_{\rm E}$, we need to obtain the inverse temperature of the environment. As indicated in the previous section and detailed in Ref.~\cite{tajik2023verification}, we have obtained the inverse temperature of the composite system with respect to the pre-quench Klein-Gordon Hamiltonian experimentally. Subsequently, we construct the thermal state corresponding to this temperature in the continuum limit simulation. We obtain the inverse temperature of the environment $\beta_{\rm E}$ for different subregion sizes, by demanding the environment's energy to match that of a Gibbs state. We then interpolate the profile for each pixel. We note that the post-quench temperature is slightly higher than the temperature measured with respect to the pre-quench Hamiltonian and $\beta_{\rm E}>0$ 
is nearly homogeneous across different subregion sizes.

To calculate the energy of a Gibbs state for the environment in the massless Klein-Gordon Hamiltonian, we begin with the partition function $Z_{\rm E}(\beta_{\rm E})=Z^{\rm zm}_{\rm E}(\beta_{\rm E})Z^{\rm ho}_{\rm E}(\beta_{\rm E})$~\cite{michel2016entanglement}.
Due to the compactified nature of the field, the zero mode partition function can be expressed in terms of the Jacobi-theta function. Given our experimental parameters, it approximates to
\begin{equation}
    Z^{\rm zm}_{\rm E}(\beta_{\rm E})=\sum_{n=-\infty}^{\infty}e^{-\frac{\beta_{\rm E}\hbar u_{\rm E}}{2}n^2}\approx \sqrt{\frac{2\pi}{\beta_{\rm E}\hbar u_{\rm E}}}\; .
\end{equation}
The harmonic oscillator partition function is
\begin{equation}
    Z_{\rm E}^{\rm ho}(\beta_{\rm E})=\prod_{k=1}^{N_{\rm E}-1}\frac{1}{2\sinh{(\beta_E\hbar\omega^{\rm E}_k/2)}}\; .
\end{equation}
The energy is then $E_{\rm E}(\beta_{\rm E})=-\partial_{\beta_{\rm E}}[\log{Z_{\rm E}(\beta_{\rm E})}]=E_{\rm E}^{\rm zm}(\beta_{\rm E})+E_{\rm E}^{\rm ho}(\beta_{\rm E})$. The zero-mode contribution to the energy is
\begin{equation}
	\begin{aligned}
		E^{\rm zm}_{\rm E} = \frac{1}{Z_{\rm 0}^{\rm E}}\sum_{n=-\infty}^{\infty}e^{-\beta_{\rm E}\frac{\hbar u_{\rm E}}{2} n^2}\left(\frac{\hbar u_{\rm E}}{2}n^2 \right)\;,
	\end{aligned}
\end{equation}
which approximates to $1/(2\beta_{\rm E})$ for our experimental parameters. The harmonic oscillator contribution to the energy is
\begin{equation}
    E_{\rm E}^{\rm ho}=\sum_{k=1}^{N_{\rm E}-1}\frac{\hbar\omega_k^{\rm E}}{2}\coth{(\beta_{\rm E}\hbar\omega_k^{\rm E}/2)}\; .
\end{equation}
Consequently, we can express then quantum relative entropy of the environment state $\varrho_{\rm E}$ with respect to a Gibbs state at inverse temperature $\beta_{\rm E}$ as
\begin{equation}
	\begin{aligned}
		D(\varrho_{\rm E}||\gamma_{\rm E}^{\beta_{\rm E}})
		=-S(\varrho_{\rm E})+\beta_{\rm E}\Tr[\varrho_{\rm E} H_{\rm E}]+\log{Z_{\rm E}(\beta_{\rm E})}
	\end{aligned}
\end{equation}
and we obtain $\Delta D$ accordingly.
%%%%%%%%%%
\clearpage
\widetext

\begin{center}
\textbf{\large Supplementary Information}
\end{center}
\setcounter{equation}{0}
\setcounter{figure}{0}
\setcounter{table}{0}
\setcounter{page}{1}
\makeatletter
\renewcommand{\theequation}{S\arabic{equation}}
\renewcommand{\thefigure}{S\arabic{figure}}

\subsection*{Unitarity of dynamics}
\begin{figure*}[h]
    \centering
        \includegraphics[width=0.99\textwidth]{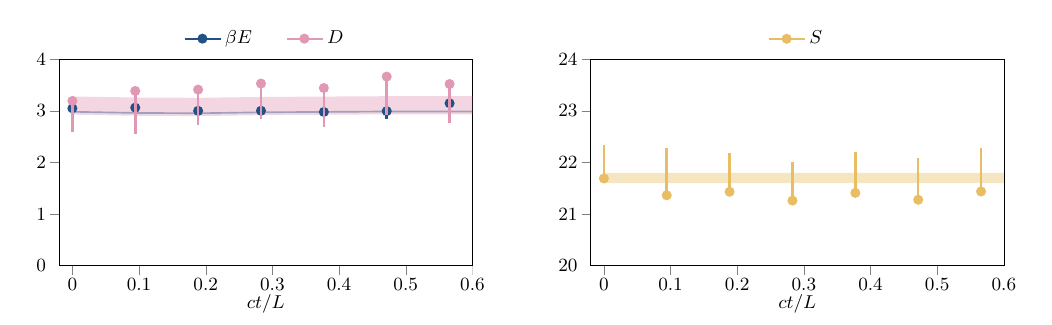}
    \caption{\textbf{Unitarity of time evolution}. The energy, relative entropy and von Neumann entropy remain globally conserved over time, providing evidence of the unitarity of the dynamics. See Fig.~\ref{fig:time_evolution_main} for more details on the error bars and shaded areas.}
    \label{fig:unitarity}
\end{figure*}

We demonstrate the global conservation of both von Neumann entropy and energy throughout the time evolution in our experiment, as shown in Fig.~\ref{fig:unitarity}. This observation provides direct evidence that the underlying dynamics are unitary, a critical assumption for the validity of using Landauer's principle for characterising the out-of-equilibrium dynamics in our experiment.

\begin{figure*}[h]
    \centering
        \includegraphics[width=0.99\textwidth]{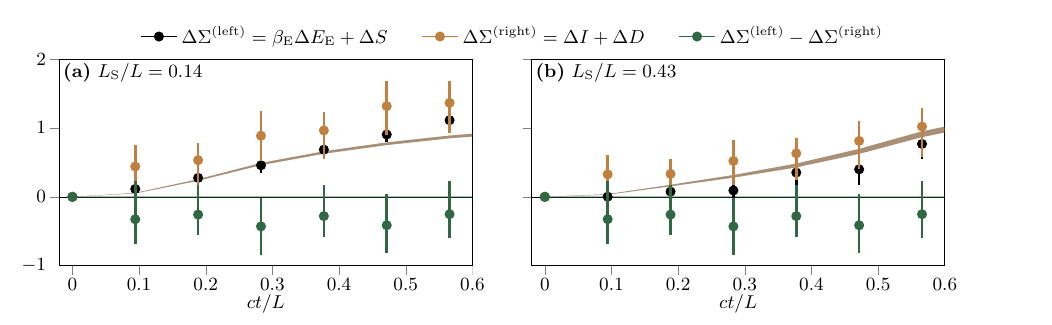}
    \caption{\textbf{Comparing the two different decompositions of generalised entropy production}. We observe good agreement between $\Delta\Sigma^{(\rm left)}$ and $\Delta\Sigma^{(\rm right)}$, as shown here as a function of time for subregion size ratios \textbf{a,} $L_{\rm S}/L=0.14$ and \textbf{b,} $L_{\rm S}/L=0.43$. This provides further evidence that the assumptions behind Landauer's principle are well satisfied in our experiment.}
    \label{fig:delta_sigma_evo_supp}
\end{figure*}

Under the assumption of global unitary dynamics for the composite system's quantum state $\varrho_{\rm SE}$, Landauer's principle (Eq.~(\ref{eq:Landauer})) can be derived. Let us define the two equivalent 
decompositions (or \textit{definitions}) of generalised entropy production, $\Delta\Sigma$, as
\begin{equation}
    \Delta\Sigma^{\rm (left)} := \beta_{\rm E}\Delta E_{\rm E}+\Delta S,\qquad \Delta\Sigma^{\rm (right)} := \Delta I+\Delta D.
\end{equation}
In our theoretical simulations, the equality $\Delta\Sigma^{\rm (left)}=\Delta\Sigma^{\rm (right)}$ is satisfied by construction, as it follows directly follows from the assumptions of unitary dynamics. In the experiment, however, this equality is far from guaranteed. Each of the quantities $\beta_{\rm E}\Delta E_{\rm E}, \Delta S, \Delta I$ and $\Delta D$ is computed independently from tomographically reconstructed covariance matrices obtained at different times $t$, where at each time step different time intervals are used for the reconstruction (Fig.~\hyperref[fig:setupandtomography]{\ref*{fig:setupandtomography}c}). Nevertheless, Fig.~\ref{fig:delta_sigma_evo_supp} shows that the two different decompositions $\Delta\Sigma^{(\rm left)}$ and $\Delta\Sigma^{(\rm right)}$ match well. This agreement thus further validates the assumptions underpinning Landauer's principle, namely, unitarity and valid density matrices (or covariance matrices, here) throughout the time evolution.

\subsection*{Scaling of absolute quantities}
\begin{figure*}[h]
    \centering
        \includegraphics[width=0.99\textwidth]{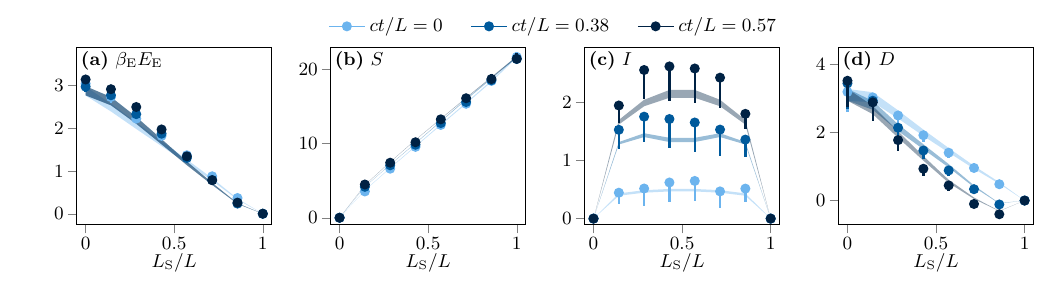}
    \caption{\textbf{Scaling of different absolute quantities in Landauer's principle}. The quantities are shown as a function of subregion size for various times $ct/L=0.19, 0.38, 0.57$. The experimental data is in agreement with the quantum field theory simulation results using Neumann boundary conditions.}
    \label{fig:scaling_absolute_supp}
\end{figure*}

We include a plot of the absolute quantities of those in Eq.~(\ref{eq:Landauer}) in Fig.~\ref{fig:scaling_absolute_supp} for the experiment.

\subsection*{Continuum limit and role of boundary conditions}
In this section we compare the effect of Neumann boundary conditions ($\partial_z\varphi(z)|_{z=0, L}=0$) and Dirichlet boundary conditions ($\varphi(z)|_{z=0, L}=0$) on the information-theoretic quantities within Landauer's principle Eq.~(\ref{eq:Landauer}) in the continuum limit. The quantum field is discretised on a  sufficiently fine grid to ensure convergence, achieved when higher momentum modes no longer affect the results significantly. The simulation parameters closely match experimental values, assuming a homogeneous density profile $n_{\rm 1D}$.

\begin{figure*}[h]
    \centering
        \includegraphics[width=0.99\textwidth]{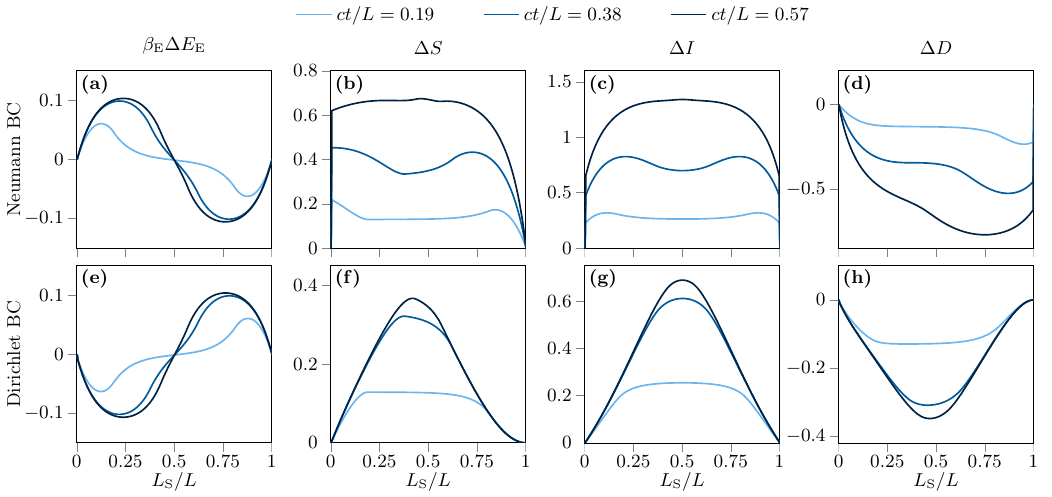}
    \caption{\textbf{Scaling of different quantities in Landauer's principle in the continuum limit for Neumann and Dirichlet boundary conditions}.  Quantum field theory simulations are used in the continuum limit, similar to Fig.~\ref{fig:scaling_main}, but without accounting for experimental inhomogeneities. \textbf{a)-d)} Neumann boundary conditions and \textbf{e)-h)} Dirichlet boundary conditions.}
    \label{fig:scaling_supp}
\end{figure*}

The plots showing the scaling with subregion size at various times are presented in Fig.~\ref{fig:scaling_supp}. The scaling of $\beta_{\rm E}\Delta E_{\rm E}$ can be explained by the initial symmetric energy density profile, which is higher at the edges than in the bulk for Neumann boundary conditions and lower for Dirichlet boundary conditions (Fig.~\hyperref[fig:DBC_main]{\ref*{fig:DBC_main}a-b}). The dominant contribution for the considered quench, $\Delta I$, shows scaling symmetrically about $L_{\rm S}/L=0.5$ by construction. For Dirichlet boundary conditions, small systems (or environments) do not increase in correlations with its environment for $ct/L_{\rm S}>1$ as correlations reach a plateau phase in time (Fig.~\hyperref[fig:DBC_main]{\ref*{fig:DBC_main}e}). Neumann boundary conditions, however, actually yield $\Delta I$ to be higher for the case of a small system than for the case of system and environment being of equal size. As discussed in the main text, the zero mode causes the results for Dirichlet boundary conditions, which are well described by the quasiparticle picture, to diverge from those for Neumann boundary conditions after $ct/L_{\rm S} > 1$.  

\subsection*{Exploiting the extremality of Gaussian states}
The validity of the Gaussian tomography scheme has been confirmed by measuring the normalised averaged connected fourth-order correlation functions, as demonstrated in previous works~\cite{schweigler2021decay, gluza2022mechanisms, tajik2023verification}. However, leveraging properties of Gaussian quantum information, we argue that the information-theoretic quantities in Landauer's principle can also be estimated using our second-moment tomography scheme, even for time evolution under a non-Gaussian model with significant higher-order correlation functions. This justifies the potential use of our experimental protocol under the lens of Landauer's principle for more complex protocols and interacting models.

To begin, we define $G$ as the Gaussian state that shares the same second moments as the \textit{true} state $\varrho$ that may be non-Gaussian, reflecting a preparation involving interactions. Central to the argument are several extremality properties of Gaussian states. Firstly, Gaussian states $G$ have the largest von Neumann entropy $S(G)\geq S(\varrho)$ among all states $\varrho$ with a given covariance matrix.
Secondly, we can make use of the property
\begin{equation}\label{eq:property}
\Tr[\varrho(t)\log{G}(0)]=\Tr[G(t)\log{G}(0)]\; ,
\end{equation}
as any Gaussian state can be written as a matrix exponent of a quadratic polynomial, thereby causing the matrix logarithm of the relative entropy to cancel out. Thirdly, Ref.~\cite{eisert2005gaussian} has demonstrated that the quantum conditional entropy is also maximised for Gaussian states, and the  inequality 
\begin{equation}
S(\varrho_{\rm S}(t))-S(\varrho_{\rm SE}(t))\geq S(G_{\rm S}(t))-S(G_{\rm SE}(t))
\end{equation}
is true. This is less obvious, as the expression under consideration is 
a difference of von Neumann entropies. 
The generalised entropy production of the true dynamics $\Delta\Sigma:=\beta_{\rm E}\Delta E_{\rm E}+\Delta S= \Delta I+\Delta D$ can be written as
\begin{equation}\label{eq:mi plus re}
\begin{aligned}
\Delta \Sigma:=\Delta I+\Delta D\\
=S(\varrho_{\rm S}(t))+S(\varrho_{\rm E}(t))-S(\varrho_{\rm SE}(t))-S(\varrho_{\rm S}(0))-S(\varrho_{\rm E}(0))+S(\varrho_{\rm SE}(0))\\-S(\varrho_{\rm E}(t))-\Tr[\varrho_{\rm E}(t)\log{\gamma_{\rm E}^{\beta_{\rm E}}}]+S(\varrho_{\rm E}(0))+\Tr[\varrho_{\rm E}(0)\log{\gamma_{\rm E}^{\beta_{\rm E}}}]\\
=S(\varrho_{\rm S}(t))-S(\varrho_{\rm S}(0))-\Tr[\varrho_{\rm E}(t)\log{\gamma_{\rm E}^{\beta_{\rm E}}}]+\Tr[\varrho_{\rm E}(0)\log{\gamma_{\rm E}^{\beta_{\rm E}}}]\\
=\Delta S-\Tr[\varrho_{\rm E}(t)\log{\gamma_{\rm E}^{\beta_{\rm E}}}]+\Tr[\varrho_{\rm E}(0)\log{\gamma_{\rm E}^{\beta_{\rm E}}}]\; ,
\end{aligned}
\end{equation}
where we have used the unitary invariance of the von Neumann entropy. Clearly, the generalised entropy production $\Delta\Sigma^G:=\beta_{\rm E}\Delta E_{\rm E}^G+\Delta S^G=\Delta I^G+\Delta D^G$ assuming an initially Gaussian state and assuming Gaussian dynamics is given by
\begin{equation}
\begin{aligned}
\Delta\Sigma^G:=\Delta I^{G}+\Delta D^{G}\\=\Delta S^G-\Tr[G_{\rm E}(t)\log{\gamma_{\rm E}^{\beta_{\rm E}}}]+\Tr[G_{\rm E}(0)\log{\gamma_{\rm E}^{\beta_{\rm E}}}],\\
\end{aligned}
\end{equation}
where $\Delta E_{\rm E}^G=\Tr[(G_{\rm E}(t)-G_{\rm E}(0))H_{\rm E}]$, $\Delta S^G=S(G_{\rm S}(t))-S(G_{\rm S}(0))$, $\Delta I^G=I^G_{\rm SE}(t)-I^G_{\rm SE}(0)$ and $\Delta D^G=D(G_{\rm E}(t)||\gamma_{\rm E}^{\beta_{\rm E}})-D(G_{\rm E}(0)||\gamma_{\rm E}^{\beta_{\rm E}})$. By utilising the property in Eq.~(\ref{eq:property}) and considering that $\gamma_{\rm E}^{\beta_{\rm E}}$ is Gaussian we can write
\begin{equation}
    \Delta \Sigma = \Delta S-\Tr[G_{\rm E}(t)\log{\gamma_{\rm E}^{\beta_{\rm E}}}]+\Tr[G_{\rm E}(0)\log{\gamma_{\rm E}^{\beta_{\rm E}}}]=\Delta S-\Delta S^G+\Sigma^G\; .
\end{equation}
This also implies that
\begin{equation}
    \Delta E_{\rm E} = \Delta E_{\rm E}^G
\end{equation}
demonstrating that covariance matrices are entirely sufficient for capturing the energetic dynamics, even when the true dynamics are not Gaussian. This is important, as in this way, one can measure second moments experimentally and still get bounds for the true state that may show signatures of interactions.

To proceed further, we must assume that the initial state post-quench is also Gaussian, i.e., $\varrho_{SE,S,E}=G_{SE,S,E}$. At this step, we have not made the additional assumption of Gaussian dynamics to bound the quantities in Landauer's principle. By the maximality of entropy for Gaussian states, we have $\Delta S\leq \Delta S^{G}$, we quickly deduce that the generalised entropy production is upper-bounded by the Gaussian value, i.e., $\Delta\Sigma \leq \Delta\Sigma^G$. Similarly, we infer that the change in mutual information is upper-bounded by the Gaussian value~\cite{eisert2005gaussian}, i.e., $\Delta I\leq \Delta I^G$. Utilising Eq.~(\ref{eq:property}), we find that the change in relative entropy is lower-bounded by the Gaussian value, i.e., 
\begin{equation}
\Delta D\geq \Delta D^G\; .
\end{equation}
However, in both cases, these inequalities alone do not provide quantitative information on the extent to which these quantities dominate each other in the true state. To gain a more precise understanding of the relative contributions of mutual information and relative entropy in this context, a more comprehensive theoretical model and advanced experimental methods are required for an experiment modelled by an arbitrary interacting theory. 

An alternative approach to probing Landauer's principle in interacting models may involve calculating classical entropies instead of their quantum counterparts. Recent work~\cite{haas2024area, deller2024area} shows that classical entropies, which can be determined without full knowledge of the composite system's state and are more accessible experimentally, can still capture key features of their quantum analogues. 
\newpage
\bibliography{bibliography}

\end{document}